\documentclass[epj]{svjour}
\usepackage{graphics}
\usepackage{amsmath}
\usepackage{amssymb}
\begin{document}

\title{Spacial and temporal dynamics of the volume fraction of the colloidal particles inside a drying sessile drop}
\subtitle{Application to human serum albumin}
\author{Yu.Yu. Tarasevich \and I.V. Vodolazskaya \and O.P. Isakova
}                     
\institute{Astrakhan State University, 20A Tatishchev Str., Astrakhan, Russia,
414056}
%

\titlerunning{Dynamics of the colloidal particles inside a drying sessile drop}

\date{Received: date / Revised version: date}
%
\abstract{ Using lubrication theory, drying processes of sessile colloidal
droplets on a solid substrate are studied. A simple model is proposed to
describe temporal dynamics both the shape of the drop and the volume fraction
of the colloidal particles inside the drop. The concentration dependence of the
viscosity is taken into account. It is shown that the final shapes of the drops
depend on both the initial volume fraction of the colloidal particles and the
capillary number. The results of our simulations are in a reasonable agreement
with the published experimental data. The computations for the drops of aqueous
solution of human serum albumin (HSA) are presented.
} 
\maketitle

\section{Introduction} \label{intro}
The patterns observed after the complete desiccation of sessile drop of a
biological fluid such as serum, cerebrospinal fluid, urine \textit{etc.}
attract the attention of the researchers at least since the
1950s~\cite{Sole1954,Koch1954,Koch1956}. In the last few decades, the interest
to the pattern formation in drying drops of the biological fluids growths
stably in the contexts of numerous applications (\textit{e.g.} medical
tests~\cite{Savina1999,Shabalin2001,Rapis2002}, drug
screening~\cite{Takhistov2002}, biostabilization~\cite{Ragoonanan2008}
\textit{etc}). State of the art can be found
in~\cite{Tarasevich2004,Yakhno2009}.

More extensive investigations are performed in the adjacent field that is
evaporative selfassembly of nano- and microparticles from an evaporating
sessile drop. This process is known as an important surface-patterning
technique which has potential application in optical, microelectronic, and
sensory devices.

Several models of deposit formation in drying drops were proposed during last
few
years~\cite{Parisse1996,Deegan2000,Fischer2002,Popov2005,Widjaja2008,Attinger2009,Zheng2009,Witten2009,Sefiane2009Langmuir,Petsi2010,Kistovich2010}.
Mostly, the particles  were considered as noninteracting and the formed deposit
hasn't any effect on both evaporation and bulk flows. It is rather clear that
these assumptions restrict the possible application of the proposed models to
the particular kind of the solutes.

One have to utilize quite different approach, if a phase transition (sol to gel
or sol to glass) occurs as a result of increasing the particle volume fraction
due to the solvent loss and internal flows. The models interpreting deposit as
impermeable  both for hydrodynamical flows and evaporation were proposed
recently~\cite{Okuzono2009,Vodolazskaya2010}.

Thus, the work~\cite{Okuzono2009} is devoted to study the drying processes of
polymer solutions on a solid substrate enclosed by bank. Drying process is
studied in the slow limit of the solvent evaporation. The model is based on
mass conservation. The master equation is a diffusion-advection equation.

Processes inside desiccated sessile drop depend on various physical and
chemical conditions such as the wetting property of the substrate and surface
roughness, the particle size, the size distribution, the particle concentration
in the droplet, the relative humidity of the evaporation environment, ionic
strength, and pH.

If the initial volume fraction of the colloidal particles inside a drop is
large enough, the contact line does not recede during the evaporation process,
depinning and stick-slip motion is not observed. In particular, it was shown
that desiccation of the colloidal suspensions considerably depends on the
competition between evaporation and gelation~\cite{Pauchard1999}.
Classification of the possible regimes of the desiccation can be done in terms
of the desiccation time $t_\text{d}$ and the gelation time $t_\text{g}$. Our
investigation is restricted to the case when the desiccation time is very short
compared with gelation time $t_\text{d} \ll t_\text{g}$. In this case  a solid
gelled or glassed foot builds up near the drop edge, while the central part
remains fluid.

For this particular situation, the desiccation process can be divided into
several stages:
\begin{description}
  \item[Pregelation stage.]  The early time regime
where a gel-like phase does not appear and the whole volume of the drop is
liquid. During this stage, the solvent evaporates, the colloidal particles are
transferred by the outward flow  to the drop edge and accumulate near the
contact line.

  \item[Gelation stage.]
The concentration of colloidal particles at the contact line reaches the
gelation concentration $C_\text{g}$ above which a gel or glass phase appears.
The gel (glass) phase region begins to develop from the edge and the phase
front monotonically move from edge to center.
  \item[Postgelation stage.] The drop is solid-like, the loss of solvent
  confined among the colloidal particles evaporates very slowly, the cracks
 appeare all around the drop.
\end{description}

The first stage was  investigated in detail in the series of
works~\cite{Tarasevich2007TP,Tarasevich2007epje,Tarasevich2010TP}, as well as
in~\cite{Okuzono2009}. In the present article, we focus efforts on the second
stage only.

The construction of this paper is as follows. In Section~\ref{sec:Model}, we
describe our model. In Section~\ref{sec:results} the obtained results are
discussed. Finally, we summarize our results and conclude this paper.

\section{Model and assumptions}
\label{sec:Model}

\subsection{Problem Formulation}\label{subsec:base}

We will consider axisymmetric sessile drop of a colloidal solution. We assume
that volume fraction of the dispersed particles is high enough to ensure strong
adhesion of the drop, in such a way the contact line of a drop is pinned during
the drying process. When the environmental conditions are homogeneous and
isotropic, the flow field inside the drop has to be axisymmetric, too.

We will use cylindrical coordinates $(r,z,\phi)$, because they are natural for
the geometry of sessile drops. The origin is chosen in the center of the drop
on the substrate. The coordinate $z$ is normal to the substrate, and the
substrate is described by $z=0$, with $z$ being positive on the droplet side of
the space. The coordinate $r$ is the polar radius. Due to the axial symmetry of
the problem and our choice of the coordinates, no quantity depends on the
azimuthal angle. In this case flow field inside a drop has only two components:
    $$
    \mathbf v (r,z,t) = u(r,z,t){\mathbf n_r} + w(r,z,t){\mathbf n_z}.
$$

The density of particles is supposed to be equal to the density of pure
solvent. This density, $\rho $, is presumed to be constant during desiccation.
These assumptions are quite reasonable for the biological fluids.

Moreover, we suppose that diffusion transfer is negligibly compared with
advection.

Only  small drops are of the practical interest, thus, we assume that the
droplet is sufficiently small so that the surface tension is dominant, and the
gravitational effects can be neglected. The temperature gradient along the
interface of the droplet is assumed to be significantly weak so that thermal
Marangoni flow is not induced. Moreover, we will neglect the concentration
dependence of the surface tension  $\sigma $ (\textit{i.e.} the concentration
Marangoni flow is absent, too).

The evaporation process can be considered as quasi-steady and we will suppose
that system is steady at every moment.

Above assumptions allow us to apply the lubrication
approximation~\cite{Burelbach1988} to the drop. This approach was used
in~\cite{Fischer2002,Okuzono2009}, too. By applying the lubrication
approximation to the continuity and Navier-Stokes equations, fluid flow is
governed by
    $$\begin{aligned}
  - \frac{\partial p}{\partial r} + \eta \frac{{{\partial ^2}u}}{{\partial {z^2}}}&= 0 ,  \\
  - \frac{\partial p}{{\partial z}}& =0,  \\
  \frac{1}{r}\frac{{\partial (ru)}}{\partial r} + \frac{{\partial w}}{{\partial z}} &= 0,  \\
\end{aligned} $$
where $p$ is the pressure and $\eta$ is the viscosity.

The boundary conditions should be written as
    \begin{equation}
    \begin{aligned}
  {\left. p \right|_{z = h}} &=- \frac{\sigma }{r}\frac{\partial }{\partial r} \left( {r\frac{\partial h}{\partial r}} \right),
&\quad
  {\left. {\frac{{\partial u}}{{\partial z}}} \right|_{z = h}}& = 0,
  \\
  {\left. u \right|_{z = 0}} &= 0,
&\quad
  {\left. w \right|_{z = 0}} &= 0,  \end{aligned}
  \label{eq:bc}
\end{equation}
where $h(r,t)$ is the droplet height profile.

In the lubrication approximation, the velocity field can be written as a
function of drop profile $h(r,t)$
\begin{equation}
    \begin{aligned}
  u &=- \frac{\sigma }{\eta }\frac{\partial }{\partial r}\left( {\frac{1}{r}\frac{\partial }{\partial r}\left( {r\frac{\partial h}{\partial r}} \right)} \right)\left( {\frac{{{z^2}}}{2}- hz} \right),\\
  w &= \frac{\sigma }{r}\frac{\partial }{\partial r}\left( {\frac{r}{\eta }\frac{\partial }{\partial r}\left( {\frac{1}{r}\frac{\partial }{\partial r}\left( {r\frac{\partial h}{\partial r}} \right)} \right)\left( {\frac{{{z^3}}}{6}- h\frac{{{z^2}}}{2}} \right)} \right).
\end{aligned}
\label{eq:uw}
\end{equation}

The height-averaged radial velocity calculated using~\eqref{eq:uw} is given by
\begin{equation}
\left\langle u \right\rangle  = \frac{1}{h}\int\limits_0^h {u\,dz}  =
\frac{{{h^2}}}{3}\frac{\sigma }{\eta }\frac{\partial }{\partial r}\left(
{\frac{1}{r}\frac{\partial }{\partial r}\left( {r\frac{\partial h}{\partial r}}
\right)} \right). \label{eq:averu}\end{equation}

From the conservation of solvent, the height-evolution equation is given by
\begin{equation}
\frac{\partial h}{{\partial t}} =- \frac{1}{r}\frac{{\partial (rh\left\langle u
\right\rangle )}}{\partial r}- \frac{J}{\rho }, \label{eq:dhdt}
\end{equation}
where $J(r,t)$ is the spatial and time-dependent solvent mass flux at the
air-liquid interface.

From the conservation of solute, the height-averaged solute concentration
$\langle C(r,t)\rangle$ is given by
    \begin{equation}
    \frac{{\partial (h\langle C \rangle)}}{{\partial t}} =- \frac{1}{r}\frac{\partial }{\partial r}\left( {r h \langle C \rangle \langle u \rangle }
    \right).\label{eq:dhCdt}
    \end{equation}

If a drop is small and surface tension dominates over gravity, the shape of the
drop is close to an equilibrium shape, \textit{i.e.}  spherical cap. Without
considerable loss of precision we can suppose that a drop has more simple shape
as a paraboloid of revolution. Nevertheless, the real shape of the drop profile
near the contact line is questionable. Note that widely used assumption about
spherical cap shape leads to singularities both in evaporation flux and flow
rate~\cite{Deegan2000}. Particularly, the precursor film of a wetting drop was
discussed in~\cite{Takhistov2002}.

The subject of our investigation is only the second stage of drying process
when colloidal particles are accumulated  at the edge of the droplet. Hence, we
can suppose that drop edge has very small thickness.

Taking into account pinning and the axial symmetry of the problem, the boundary
and initial conditions for the set of equations~\eqref{eq:dhdt},
\eqref{eq:dhCdt} can be written as
    \begin{equation}\begin{aligned}
  {\left. h \right|_{t = 0}} &= h_\text{f}+{h_0}\left( {1- {{\left( {\frac{r}{R}} \right)}^2}} \right), \\
  {\left. { \langle u \rangle } \right|_{r = 0}} &= {\left. { \langle u \rangle } \right|_{r = R}} = 0,\\
  {\left. {\frac{\partial h}{\partial r}} \right|_{r = 0}} &= 0, \\ {\left. h \right|_{r = R}} &= h_\text{f}, \\
 \langle C(r, 0) \rangle &= f(r), \\ {\left. {\frac{{\partial \langle C \rangle}}{\partial r}} \right|_{r = 0}} &= 0 ,\\
\end{aligned} \label{eq:mastersys}
\end{equation}
where  ${f(r)}$ is the initial concentration of the colloidal particle,  $h_0 +
h_\text{f}$ is the initial height of the drop apex, $h_\text{f}$ is the initial
height of the drop edge ($h_\text{f} \ll h_0$).

To nondimensionalize the equations, the vertical coordinate was scaled by the
initial droplet height ${h_0}$, the radial coordinate was scaled by the drop
radius $R$, the velocities were scaled by the characteristic viscous velocity
${u_c} = \frac{{{\eta _0}}}{{\rho {h_0}}}$, where $\eta_0$ is the viscosity of
pure solvent. Time was scaled by $\frac{R}{{{u_c}}}$, vapor flux was scaled by
${J_c} = \frac{{k\Delta T}}{{L{h_0}}}$, where $k$ is the thermal conductivity
of the liquid, $\Delta T$ is the temperature difference between the substrate
and saturation temperatures, $L$ is the heat of vaporization. The
height-averaged concentration was scaled by gelation concentration
$C_\text{g}$.

Upon scaling, the evaporation number, $\mathrm{E} = \frac{{k\Delta
T}}{{\varepsilon {\eta _0}L}}$, and the capillary number, $\mathrm{Ca} =
\frac{{{\eta _0}{u_c}}}{{{\varepsilon ^3}\sigma }}$, where $\varepsilon  =
\frac{{{h_0}}}{R}$, appear as the dimensionless parameters.

In the dimensionless form, dynamics of the drop profile and the concentration
of the colloidal particles are governed by
\begin{equation}
\frac{\partial h}{{\partial t}} =- \frac{1}{r}\frac{{\partial (rh\left\langle u
\right\rangle )}}{\partial r}- J\mathrm{E}, \label{eq:dhdtdl}
\end{equation}

    \begin{equation}
    \frac{{\partial (h \langle C \rangle )}}{{\partial t}} =- \frac{1}{r}\frac{\partial }{\partial r}\left( {r h  \langle C \rangle\langle u \rangle
    }
    \right),\label{eq:dhCdtdl}
    \end{equation}
where
\begin{equation}
\left\langle u \right\rangle  = \frac{1}{\mathrm{3Ca}}\frac{h^2}{\eta
}\frac{\partial }{\partial r}\left( {\frac{1}{r}\frac{\partial }{\partial
r}\left( {r\frac{\partial h}{\partial r}} \right)} \right).
\label{eq:averudl}\end{equation} To simplify notation, we don't use any special
symbols for the dimensionless quantities.

In our consideration, we take into account concentration dependence of solvent
viscosity $\eta(r,t)$. Thus, viscosity is not a constant but varies in space
and time.

The master equations~\eqref{eq:dhdtdl} and~\eqref{eq:dhCdtdl} may be
reorganized and rewritten using the matrix notation
\begin{equation}
\label{eq:equationsSet2} \frac{\partial \mathbf{F}}{\partial t}+ \langle
u\rangle\frac{\partial \mathbf{F}}{\partial r}=\mathbf{f},
\end{equation}

\begin{equation}
\label{eq:vectorU} \mathbf{F}=
\begin{pmatrix}
h\\ \langle C \rangle
\end{pmatrix},
\end{equation}

\begin{equation}
\label{eq:vectorF} \mathbf{f}=
\begin{pmatrix}
-\dfrac{h}{r} \dfrac{\partial }{\partial r}(r\langle u\rangle) - \mathrm{E}J
\\
\mathrm{E}\dfrac{\langle C \rangle J}{h}
\end{pmatrix}.
\end{equation}

In the framework of the lubrication approximation, the evaporative mass flux
must be equal to zero at the contact line. To our best knowledge the
experimental data on evaporative flux over the free surface of a colloidal
droplet with moving phase front inside it are not published yet. We will
exploit the mass flux of evaporating liquid  derived from the heat transfer
analysis~\cite{Anderson1995}
$$
    J(r,t) = \frac {1 }{\kappa + h(r,t) },
$$
where $\kappa$~is a dimensionless nonequilibrium parameter. $\kappa \to \infty$
for nonvolatile substances and  $\kappa \to 0$ for high volatile liquids. A
conjecture that colloidal particles accumulated at the edge of the droplet must
hinder the evaporation at the contact line was proposed by
Fischer~\cite{Fischer2002}. The author introduced an additional factor in
evaporative flux so that the flux decreases exponentially near the contact
line. We suppose that the density of vapor vanishes not at the edge of a drop
but at the sol-gel phase boundary. In the other words, the vapor flux has to be
zero if the concentration of the colloidal particles reaches a critical value
($C = 1$), which corresponds to a phase transition (sol to gel or sol to
glass):
    \begin{equation}
    J(r,t) = \frac {1 -  \langle C \rangle^2 ( r,t ) }{\kappa + h(r,t) }.
    \label{eq:J}
    \end{equation}

We should notice that the results of computations depend drastically on
particular form of the evaporation flux. Our choice~\eqref{eq:J} provides
linear decreasing with time of the drop mass~\cite{Deegan2000,Annarelli2001}
and the reasonable dynamics of drop shape (see Section~\ref{sec:results}).

\subsection{Parameters of the model}\label{subsec:parameters}

The computations were performed with $\mathrm{Ca} = 0.01, 0.1, 1, 10 $ and
$\mathrm{E} = 0.01, 0.1, 1$, as well as with the $\mathrm{Ca}$ and
$\mathrm{E}$, which correspond to the drops of biological fluid used in the
medical tests. In the medical tests, the typical drop volume varies between 10
and 20~$\mu$l with diameter 5--7~mm~\cite[p.~100]{Shabalin2001}. It means that
typical drop height is $h_0 \approx 1$~mm. In general, the drop height should
be lower, since we consider only the second stage of desiccation,  when the
volume fraction of the colloidal particles reaches the critical value of
$\Phi_\text{g}$ near the drop edge. Nevertheless, the
experiments~\cite{Yakhno2004} as well as the
simulations~\cite{Okuzono2009,Tarasevich2010TP} show that the first stage lasts
only about 10~\% of desiccation time. Taking into account linear decreasing of
the drop height, we can ignore the height change. We take the middle value $R =
3$~mm as the typical radius, so $\varepsilon \approx 0.33$. Assuming $\rho =
10^3$~kg/m$^3$,  $\eta_0 = 10^{-3}$~Pa$\cdot$sec,   and $\sigma = 73 \cdot
10^{-3}$~N/m, the characteristic viscous velocity will be $u_\text{c} =
10^{-3}$~m/sec. In this case the capillary number will be $\mathrm{Ca} \approx
4 \cdot 10^{-4}$. It should be noted that $\varepsilon^2 \approx 0.1$, hence
the lubrication approximation is still valid. The rest of the quantities of use
 are  $k = 0.6$~ W/(m K), $\Delta T = 80$~K$, L = 2.25 \cdot
10^6$~J/kg, hence $\mathrm{E} \approx 3 \cdot 10^{-2}$.

The experimental data of solution viscosities of HSA for concentrations ranges
from $8.2$ up to $369$ kg/m$^3$ at pH$=7.0$ and temperature $20^\circ$~C were
taken from~\cite{Monkos2004}. The volume fraction of colloidal particles can be
calculated as
$$
\Phi= \frac{V C}{M_\text{h}},
$$
where $V = \frac{4}{3} \pi p b^3$ is the hydrodynamic volume of one dissolved
protein ($a$ and $b$ are the effective semi-axes of hydrated HSA and $p$ is the
axial ratio), $M_\text{h} = 91.675$ kDa is the molecular mass of the dissolved
protein~\cite{Monkos2004}.

\begin{figure}
\resizebox{\columnwidth}{!}{%
 \includegraphics{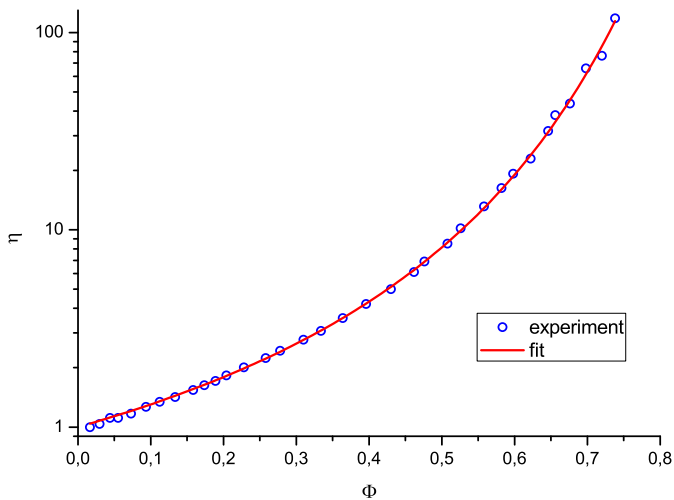}
} \caption{ Nondimensional viscosity \textit{vs.}  normalized volume fraction:
experiment~\cite{Monkos2004} and fit by Mooney~\eqref{eq:Mooney}.}
\label{fig:viscosity}
\end{figure}

Taking into account the data published in~\cite{Monkos2005}, $a=8.2$~nm, $b=
2.1$~nm, $p=3.95$, we can get the relation $\Phi = 10^{-3}C$. Then the volume
fraction was divided by $\Phi_\text{g} = 0.5$. Nondimensional data were fitted
by Mooney's formula using least square method
\begin{equation}\label{eq:Mooney}
    \eta = \eta_0 \exp\left( \frac{S\Phi}{1- K\Phi}\right),
\end{equation}
where $\eta_0$ is the viscosity of pure solvent (see fig.~\ref{fig:viscosity}).

For the normalized by $\Phi_\text{g}$ volume fraction and nondimensional
viscosity, the fitting parameters are $S = 2.42 \pm 0.07$ and $K = 0.84 \pm
0.02$.

One can calculate the density of the hydrated protein $\rho \approx
993$~kg/m$^3$, hence our assumption (Section~\ref{subsec:base}) about equality
of the solute and solvent densities holds for aqueous solution.

The initial concentration can be written as
$$
f(r) = 2 - \langle C_0 \rangle + 2 \frac { \langle C_0 \rangle - 1}{1 + \exp(
w( r-1 ))},
$$
where $w$ is an adjustable constant, it is related to the length over which the
concentration of colloidal particles increases rapidly.  $w = 10$ in all
computation presented in this article. This choice gives us a good agreement
with the calculated distribution at the end of the first stage of
desiccation~\cite{Tarasevich2007TP,Tarasevich2007epje,Tarasevich2010TP}.
Notice, that $f(0) \approx \langle C_0 \rangle$ and $f(1) = 1$.

Computations in~\cite{Tarasevich2010TP} demonstrate, that at the begin of the
second stage, the concentration in the drop center is approximately equal to
the initial concentration. Notice, that in~\cite{Okuzono2009} the concentration
in the drop center growths at the rate one and a half times. This fact is the
natural consequence of utilized uniform vapor flux over the free surface.

 $\kappa = 0.01$ and $h_\text{f} = 0.01$ in all presented simulations. Our
computations say that the results are almost insensitive to $h_\text{f}$, if
the relation $h_\text{f} \ll h_0$ is valid.

\section{Results}\label{sec:results}
\begin{figure}
\resizebox{0.8\columnwidth}{!}{%
  \includegraphics{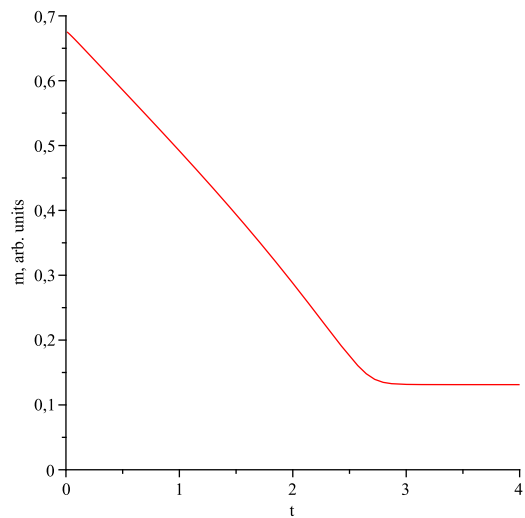}
} \caption{Temporal dynamics of mass, with $\mathrm{Ca} = 0.1$, $\mathrm{E} =
0.1$, $\Phi_0 = 0.2$. }\label{fig:massa}
\end{figure}

\begin{figure}
\resizebox{0.8\columnwidth}{!}{%
  \includegraphics{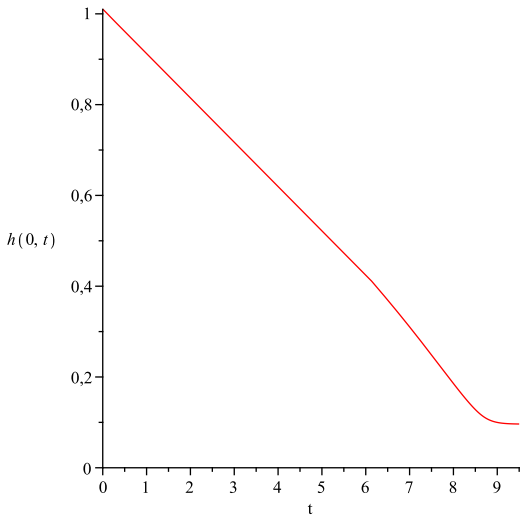}
} \caption{Temporal height dynamics, with $\mathrm{Ca} = 4 \cdot 10^{-4}$,
$\mathrm{E} = 0.03$, $\Phi_0 = 0.2$. }\label{fig:h0}
\end{figure}

\begin{figure*}
a)\resizebox{0.3\textwidth}{!}{%
\includegraphics{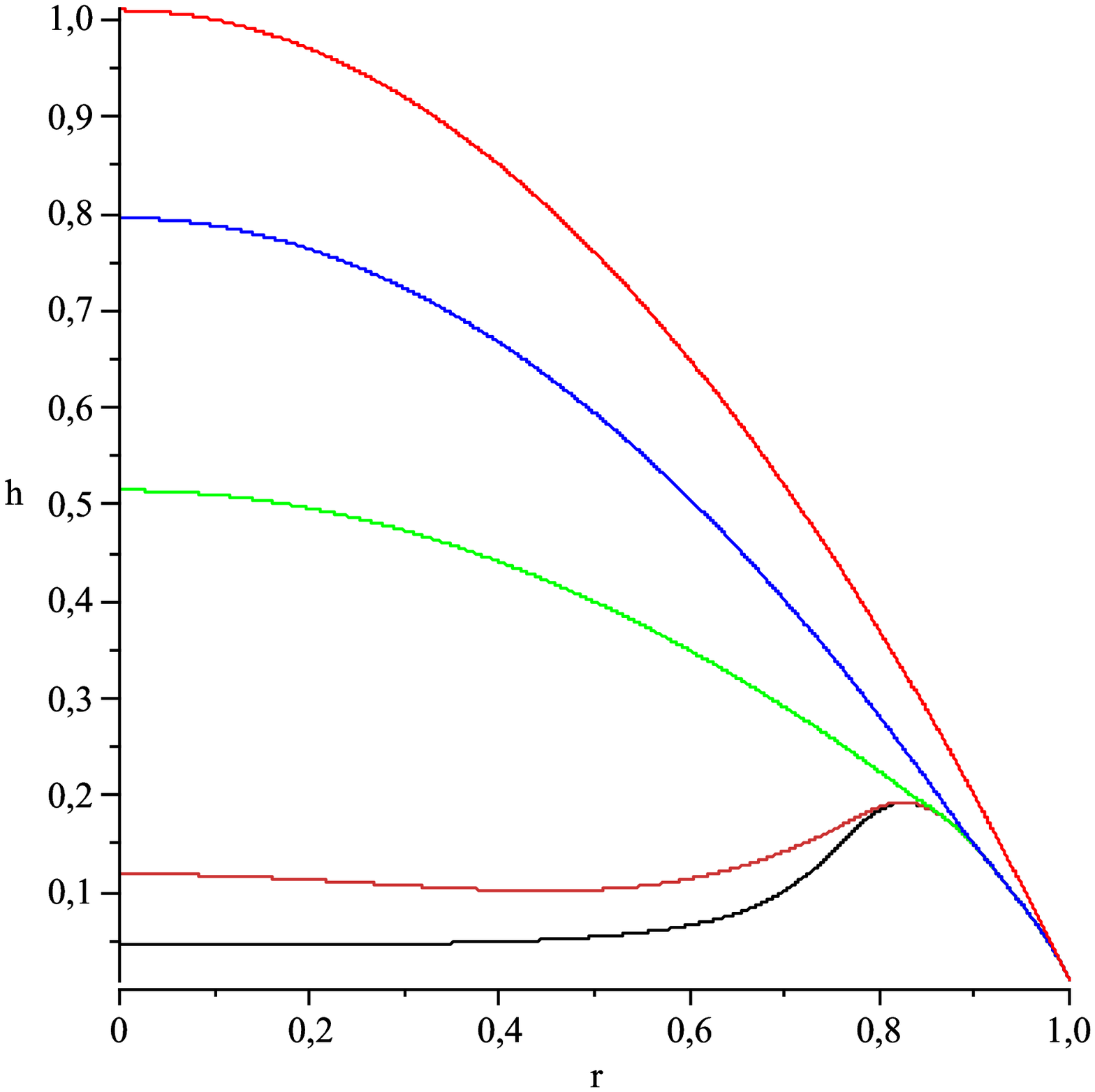}
}
b)\resizebox{0.3\textwidth}{!}{%
\includegraphics{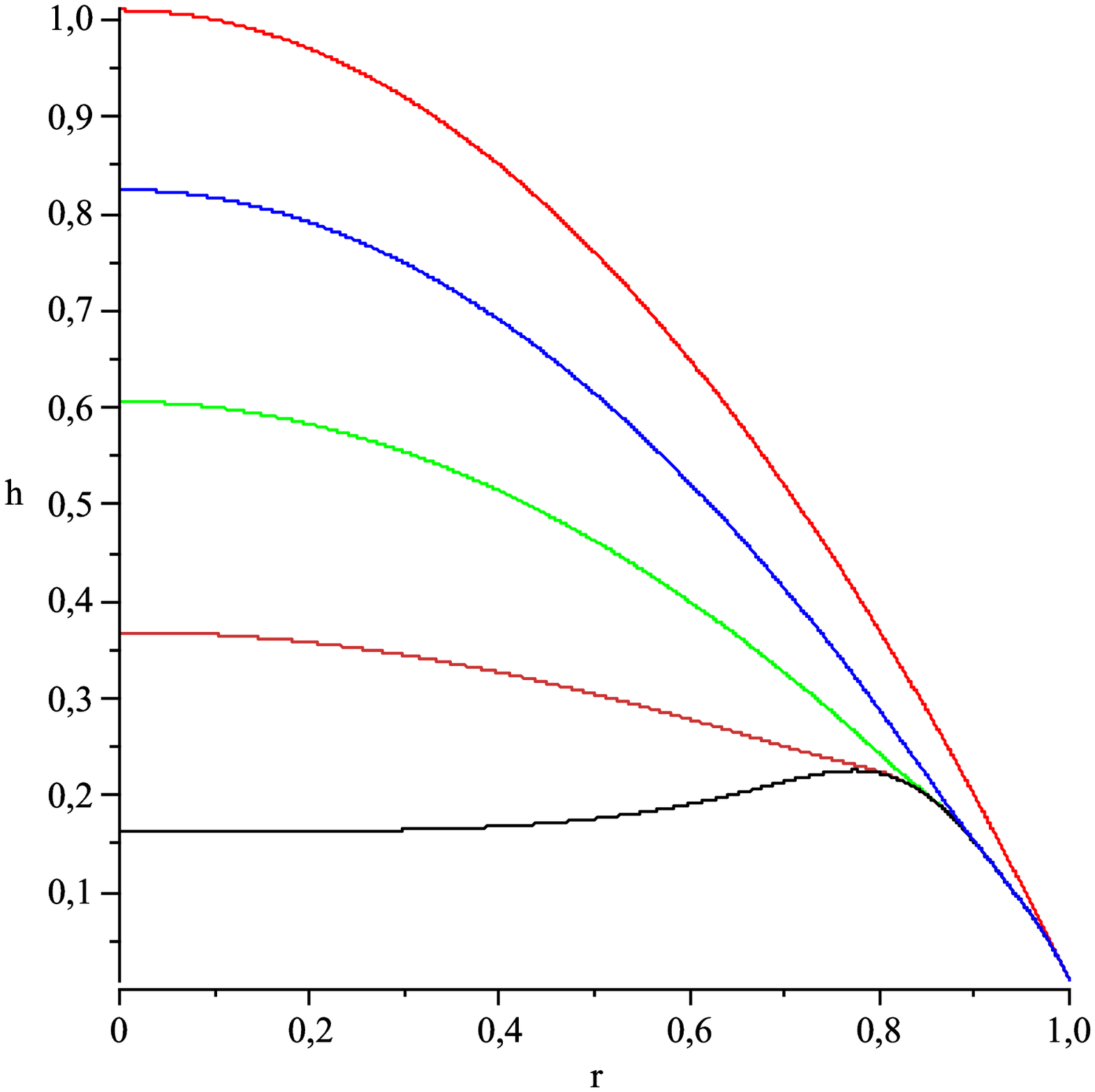}
}
c)\resizebox{0.3\textwidth}{!}{%
\includegraphics{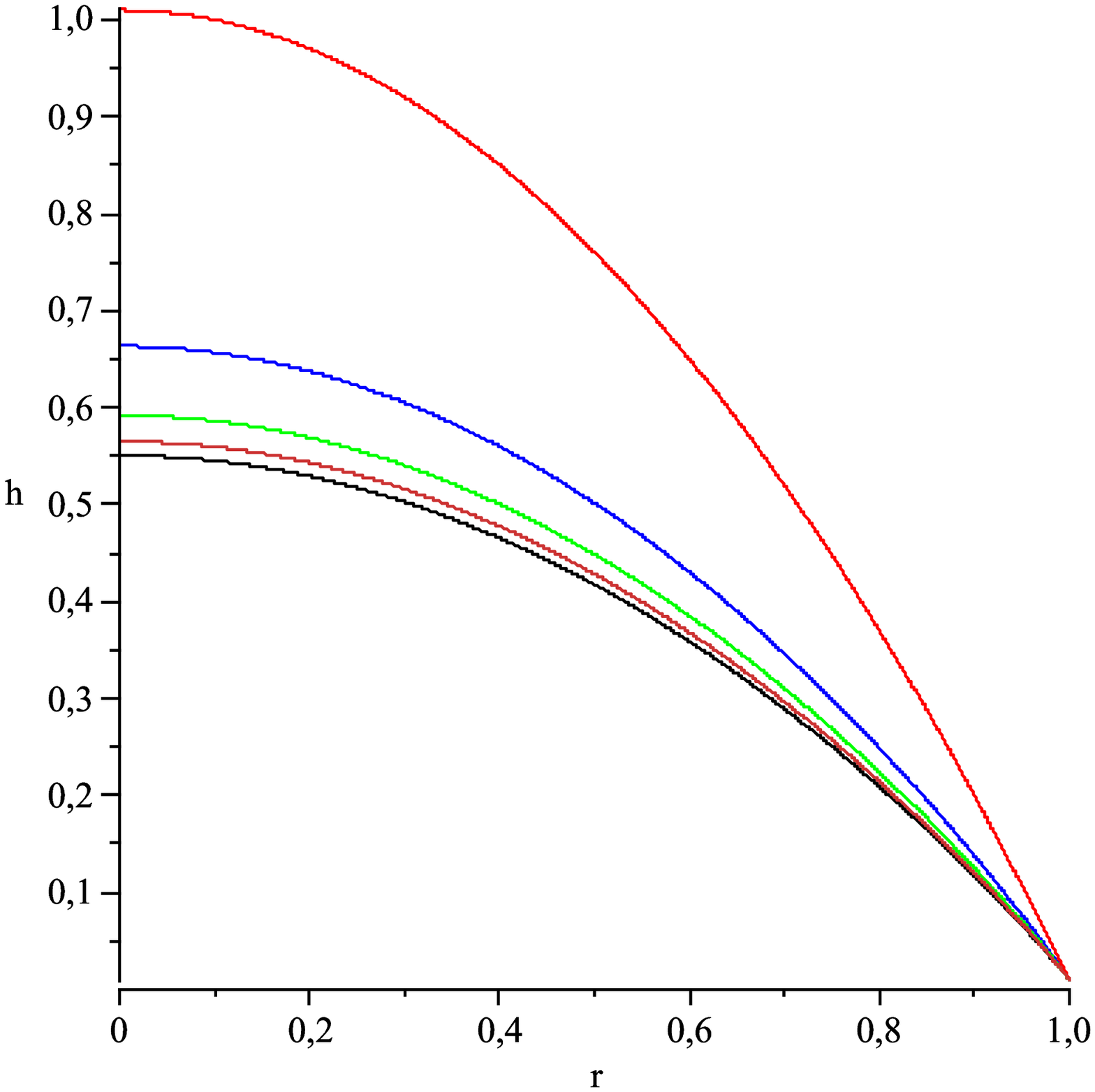}
}
 \caption{(Color online) Time evolution of the droplet height profile for $t/t_\text{max} = 0, 0.25, 0.5,
0.75, 1$, with $\mathrm{Ca} = 0.1$, $\mathrm{E} = 0.1$. a) $\Phi_0 = 0.1$, b)
$\Phi_0 = 0.25$, c) $\Phi_0 = 0.4$.  }
\label{fig:differentPhi}       
\end{figure*}

\begin{figure*}
a)\resizebox{0.3\textwidth}{!}{%
\includegraphics{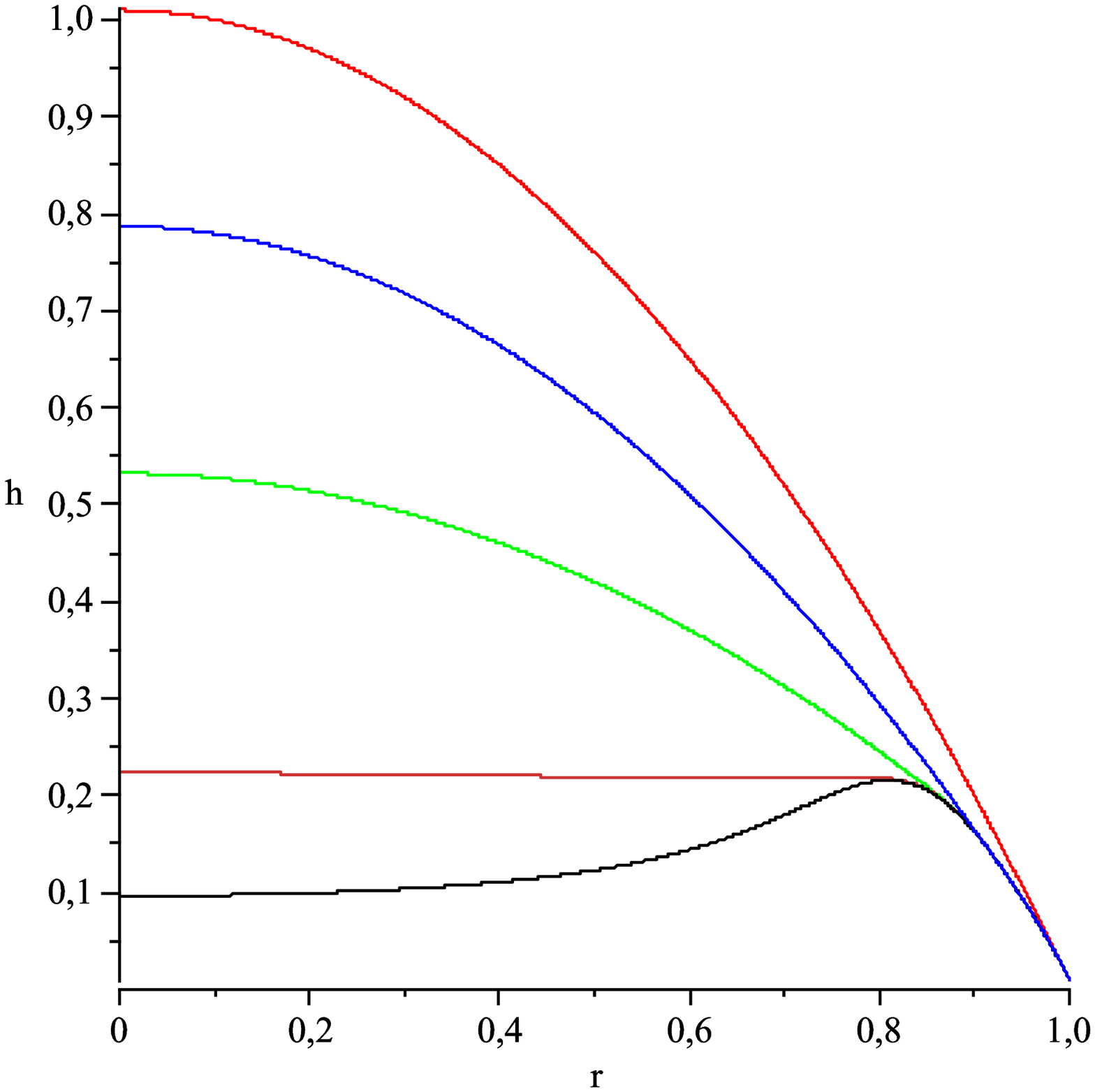}
}
b)\resizebox{0.3\textwidth}{!}{%
\includegraphics{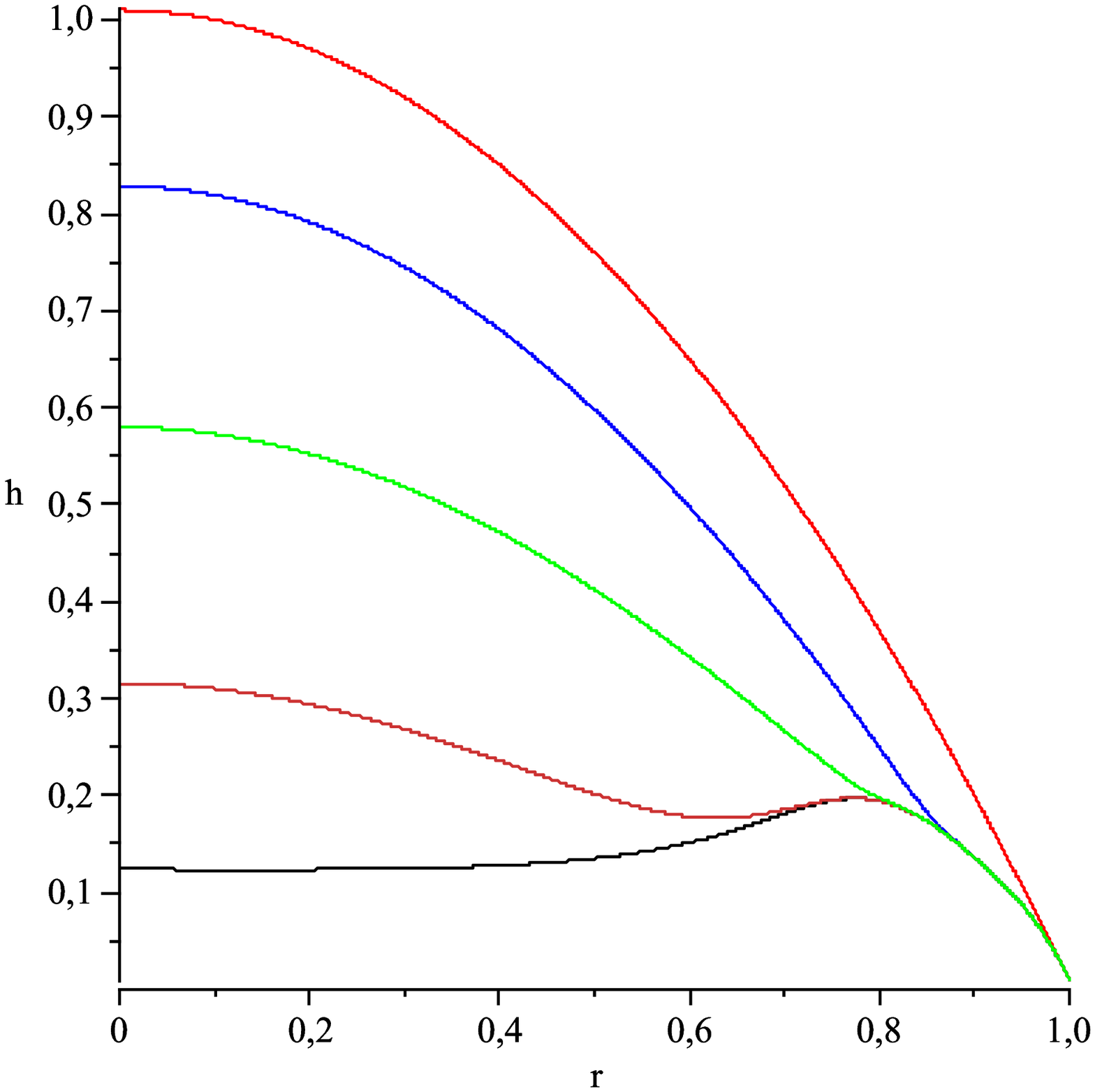}
}
c)\resizebox{0.3\textwidth}{!}{%
\includegraphics{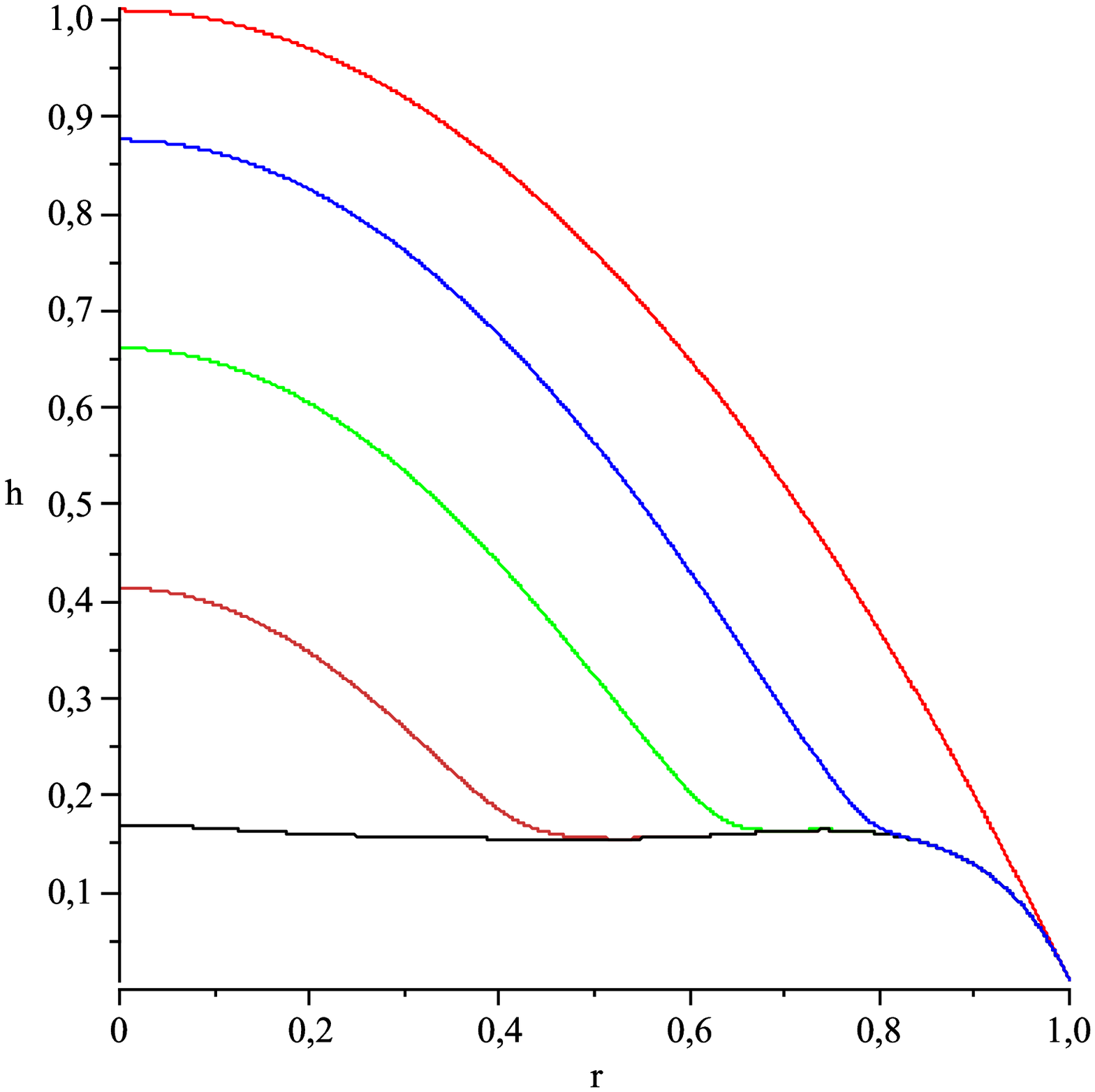}
} \caption{(Color online) Time evolution of the droplet height profile for
$t/t_\text{max} = 0, 0.25, 0.5, 0.75, 1$, with $\Phi_0 = 0.2$, $\mathrm{E} =
0.1$. a) $\mathrm{Ca} = 0.01$, b) $\mathrm{Ca} = 1$, c) $\mathrm{Ca} = 10$.  }
\label{fig:differentCa}
\end{figure*}

\begin{figure}
\resizebox{0.8\columnwidth}{!}{%
  \includegraphics{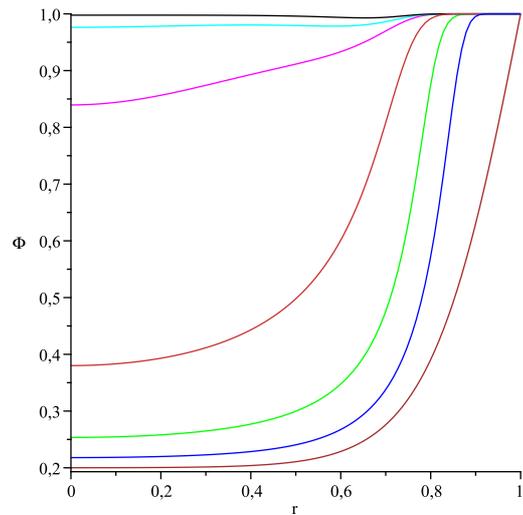}
} \caption{(Color online) Evolution of the particle volume fraction for
$t/t_\text{max} = 0, 0.25, 0.5, 0.75, 0.9, 0.95, 1$, with $\mathrm{Ca} = 0.1$,
$\mathrm{E} = 0.1$, $\Phi_0 = 0.2$.} \label{fig:concentr}
\end{figure}

\begin{figure}
\centering
\resizebox{0.8\columnwidth}{!}{%
  \includegraphics{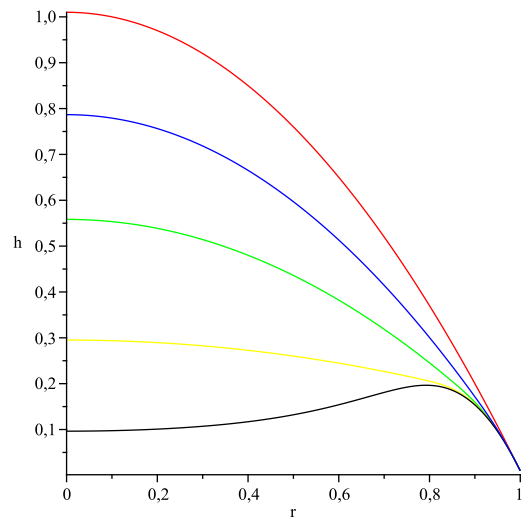}
} \caption{(Color online) Time evolution of the droplet height profile for
$t/t_\text{max} = 0, 0.25, 0.5, 0.75, 1$, with $\mathrm{Ca} = 4 \cdot 10^{-4}$,
$\mathrm{E} = 0.03$, $\Phi_0 = 0.2$. }\label{fig:profile}
\end{figure}

The results of our simulations are presented in figs.~\ref{fig:massa},
\ref{fig:h0}, \ref{fig:differentPhi}, \ref{fig:differentCa},
\ref{fig:concentr}, \ref{fig:profile}. $t_\text{max}$ occurring in the captions
relates to the time of the whole drop bulk becoming solid-like and the drop
height almost not decreasing.

Mass of the drop decreases linearly until almost all solvent lost from the
drop, then it stays nearly constant (Fig.~\ref{fig:massa}). This behavior
agrees with the published experimental results (see
\textit{e.g.}~\cite{Deegan2000,Annarelli2001}) and demonstrates that the
proposed evaporation flux~\eqref{eq:J} is rather adequate. In our simulations,
difference between mass of the completely dried drop and the initial total mass
of the colloidal particles does not exceed 1.5~\%. This fact confirms that
computations are quite correct.

Drop height decreases approximately in linearly manner until almost all solvent
evaporates from the drop, then it remains unchanged (Fig.~\ref{fig:h0}). This
behavior is typical just for the case when $t_\text{d} \ll t_\text{g}$
\cite{Pauchard1999}, which is the subject of our investigations.

We have investigated the effect of initial volume fraction of the colloidal
particles inside the drop on time evolution of the droplet height profile as
well as distribution of the colloidal particles (Fig.~\ref{fig:differentPhi}).
If the initial volume fraction is low there will be visible roller around the
edge of dried sample. If the initial volume fraction has an intermediate value
the final shape of the sample looks like a pancake, with a depressed central
zone. Indeed, if the initial volume fraction is close to the value
$\Phi_\text{g}$ the drop shape varies very low. Notice that profile evolution
dynamics during early time is very similar to the dynamics of the constant base
model proposed in~\cite{Parisse1996}.

Fig.~\ref{fig:differentCa} shows our results obtained for fixed initial volume
fraction and evaporation number and with different values of capillary number.
If capillary number is small (\textit{i.e.} surface-tension forces dominate
viscous forces) early time evolution of the droplet height profile is similar
to constant base model~\cite{Parisse1996}. If the capillary number is large,
then viscous forces will dominate surface tension forces. In this case profile
dynamics corresponds to constant angle model~\cite{Parisse1996}. It looks like
if the capillary number is intermediate, then the evolution of profile is
closer to the predictions of model proposed by Popov~\cite{Popov2005}. Thus,
the known models are suitable and correct for different particular cases and
can be reproduced within our model.

Evolution of the colloidal particle distribution is depicted in
Fig.~\ref{fig:concentr}. This result is in qualitative agreement
with~\cite{Okuzono2009}.

\begin{figure}
\centering
\resizebox{0.8\columnwidth}{!}{%
  \includegraphics{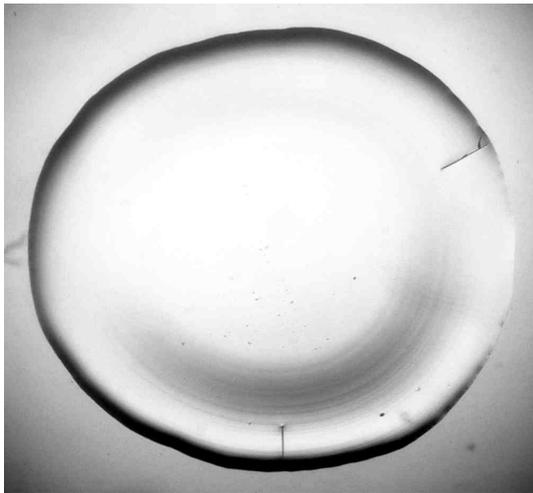}
} \caption{Sample of a drop after 16 min it was placed onto glass slide (10 w\%
of BSA in normal saline solution). (Courtesy T.A.~Yakhno).}\label{fig:drop}
\end{figure}

Moreover, some computations have been performed for the parameters which are
close to the values for samples used in the medical tests. Time evolution of
the droplet height profile is presented in Fig.~\ref{fig:profile}. Notice that
final shape of the drop is a flat pancake, with a depressed central zone
(Fig.\ref{fig:profile}) and looks like the real sample (Fig.~\ref{fig:drop}).

\section{Conclusion\label{sec:conclusion}}

The main open question of our work is related to the real form of evaporation
flux and adequacy of our choice~\eqref{eq:J}. In our opinion,  this question
can be resolved only experimentally.

The biological fluids contain both colloidal particles (proteins) and the
salts. Salts have rather high diffusivity, hence they cannot be described
using~\eqref{eq:equationsSet2}. It is of interest and of value to include in
the model additional spices with high diffusivity. The natural way to improve
the proposed model is to supplement the set of
equations~\eqref{eq:equationsSet2} with additional advection-diffusion equation
describing spatial and temporal dynamics of
salt~\cite{Tarasevich2007TP,Tarasevich2007epje,Tarasevich2010TP}.

\begin{acknowledgement}

The authors would like to thank T.A.~Yakhno for the photo
(fig.~\ref{fig:drop}). This work was supported by the Russian Foundation for
Basic Research, project no. 09-08-97010-r\_povolzhje\_a.
\end{acknowledgement}

 \bibliographystyle{epj}
 \bibliography{drops,dropsown}

\end{document}